\documentclass[useAMS,usenatbib,usegraphicx]{pasa}

\title[The Canis Major Dwarf Galaxy]{
The Canis Major Dwarf Galaxy}

\author[Lewis et al.]{
Geraint F. Lewis$^{1,5}$,
Rodrigo A. Ibata$^2$,
Michael J. Irwin$^3$, \\
Nicolas F. Martin$^2$,
Michele Bellazzini$^4$ \&
Blair Conn$^1$
\affil{
$^1$Institute of Astronomy, School of Physics, A29, 
University of Sydney, NSW 2006, Australia \\
$^2$Observatoire de Strasbourg, 11 Rue de l'Universite, 
F-6700 Strasbourg, France\\
$^3$Institute of Astronomy, Madingley Road, Cambridge, CB3 0HA, U.K.\\
$^4$INAF - Osservatorio Astronomico di Bologna, Via Ranzani 1, 40127, Bologna, Italy\\
$^5$Email: {\tt gfl@physics.usyd.edu.au}
}}
\begin{document}

\maketitle

\label{firstpage}

\begin{abstract}
Recent  observational  evidence suggests  that  the Sagittarius  dwarf
galaxy  represents  the only  major  ongoing  accretion  event in  the
Galactic  halo,   accounting  for  the  majority   of  stellar  debris
identified  there.   This paper  summarizes  the  recent discovery  of
another  potential Milky Way  accretion event,  the Canis  Major dwarf
galaxy.  This  dwarf satellite galaxy is  found to lie  just below the
Galactic  plane and  appears to  be  on an  equatorial orbit.   Unlike
Sagittarius, which is contributing  to the Galactic halo, the location
and  eventual demise  of Canis  Major  suggests that  it represents  a
building block of the thick disk.
\end{abstract}


\section{Introduction}\label{introduction}
${\rm \Lambda}$CDM  represents the current paradigm  for the formation
and evolution of  structure in the Universe. While  successul on large
scales,  focus  has recently  turned  to  its  inability to  correctly
predict   the  number   of   satellite  systems   in  galactic   halos
\citep{1999ApJ...522...82K}. This  {\it missing satellite  problem} is
apparent with the Milky  Way, with recent observations indicating that
the Milky Way has undergone  a single large accretion, the Sagittarius
Dwarf            galaxy,             in            the            last
$\sim7$Gyrs~\citep{2002MNRAS.332..921I,2003ApJ...599.1082M},   although
some        older       accretions        are        apparent       in
phase-space~\citep{1999Natur.402...53H,2003ApJ...585L.125B}. 

\begin{figure*}
\includegraphics[height=4.4in]{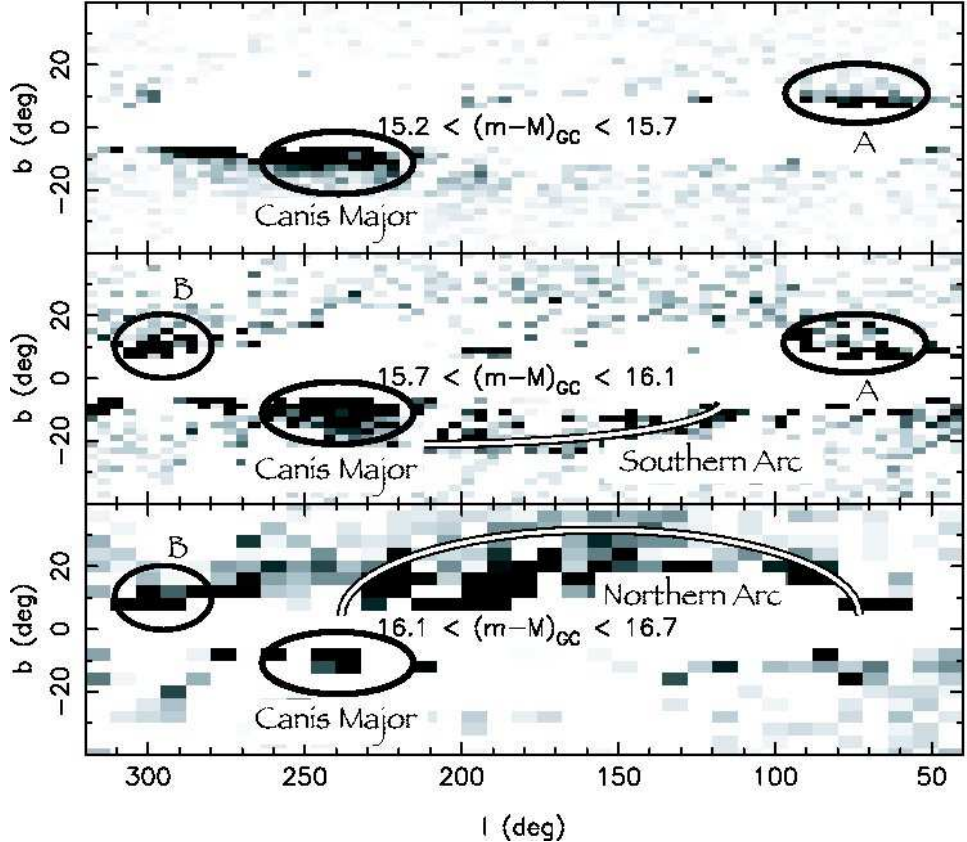}
\caption{\label{fig1} Binned M-giant  counts about the Galactic plane.
The  three  panels denote  different  magnitude  (and hence  distance)
ranges,  with the  galactocentric distance  modulus labeled  upon each
panel. Also in each panel,  the counts represent the asymmetry between
the north and south (ie -  southern data subtracted from the north and
vice-versa). The main body of  Canis Major and prominent structures in
the    M-giant    distribution    are    noted    (Figure    5    from
\citet{2004MNRAS.348...12M}).}
\end{figure*}

To  compare  with  predictions  from ${\rm  \Lambda}$CDM  models,  the
accretion history  of the Milky Way  needs to be mapped  out.  As with
many studies of the Universe,  however, our view of the Galactic halo,
and any accreting  systems, is obscured by the disk  of the Milky Way.
Recent large scale  surveys of this region, however,  have uncovered a
curious overdensity of  stars that has been interpreted  as being part
of a current accretion event that  is taking place within the plane of
the Milky Way. This paper reviews the discovery of this object and our
current understanding of its properties.

\section{The ring around the Galaxy}\label{ring}
The first  indication of  an additional halo  population of  stars was
found by ~\citet{2002ApJ...569..245N} while examining halo stars drawn
from  the  Sloan  Digital  Sky  Survey.  Taken in  a  narrow  band  of
$\sim2.5^o$  about the  celestial equator,  and selected  to  have the
colours  of F-stars, these  data revealed  a prominent  overdensity of
halo stars  which has been  interpreted as the survey  slicing through
the stream  of tidal  debris from the  Sagittarius Dwarf  Galaxy.  One
additional prominent  overdensity of  stars was identified  within the
halo, towards the Galactic anti-centre, in the direction of Monoceros,
at a galactocentric distance  of $\sim18$kpc and width $<6$kpc.  While
a   Galactic   origin   could   not   be   conclusively   ruled   out,
\citet{2002ApJ...569..245N}  suggested  that  this too  represented  a
tidal stream, but of a yet unknown disrupting companion galaxy,

Spurred  by this  discovery, \citet{2003MNRAS.340L..21I}  searched for
the signature  of the  Monoceros stream of  stars in the  Isaac Newton
Telescope   Wide    Field   Survey   data    archive.    Analysis   of
colour-magnitude  diagrams, \citet{2003MNRAS.340L..21I}  confirmed the
identification of  the distinct stellar population in  the vicinity of
Monoceros. This study also  identified the Monoceros stream population
in a number of additional  fields, revealing this population to extend
$\sim100^o$ over  the sky, within $\sim30^o$ of  the Galactic equator,
and  \citet{2003MNRAS.340L..21I}  suggest  that  the  Monoceros  stream
actually rings the  Galaxy. Fitting the main sequence  of this stellar
population in  each field, this  study estimates that the  distance to
the  stream ranges  from  $\sim15$ to  $\sim20$kpc,  with an  apparent
scale-height  of $\sim0.75$kpc.  While  they consider  the possibility
that this stellar population was an accreting dwarf, they also pointed
out that their  data is consistent with other  hypotheses including an
outer spiral  arm or unknown  flare/warp generated via a  resonance in
the Galactic disk.

At   the  same  time   as  the   \citet{2003MNRAS.340L..21I}  results,
\citet{2003ApJ...588..824Y}  presented  a   kinematic  analysis  of  a
Galactic  halo  stars  drawn  from  the SDSS.   Focusing  upon  F-star
candidates, this  study obtained spectra in  several regions, allowing
the determination of their  kinemtatic properties.  Accounting for the
Galactic contribution,  the Monoceros population  was found to  have a
velocity dispersion of $\sim25-30$km/s.  While of similar order to the
tidal debris torn from the Sagittarius dwarf, this velocity dispersion
is quite distinct  from the spheroid, thick disk or  any known warp or
flare. From these  velocities, \citet{2003ApJ...588..824Y} deduce that
the orbital velocity of the  stream of stars is prograde and (assuming
circular orbits) is $215\pm25$km/s, [An erratum to the original result
of $110\pm25$kms/s was  presented in \citet{2004ApJ...605..575Y}], and
that    the    stars    appear    to    be    relatively    metal-poor
$([\frac{Fe}{H}]=-1.6)$.    \citet{2003ApJ...588..824Y}   conclude  by
proposing  a simple  model for  the Monoceros  stream as  a disrupting
dwarf, orbiting the Milky Way  at a distance of $\sim18$kpc; the dwarf
galaxy's stars are  undulating above and below the  plane of the Milky
Way by $\sim6$kpc.

Several   additional  programs   have  focused   upon   the  Monoceros
stream~\footnote{The stream of stars  has acquired several names since
its  discovery. For  the  sake of  consistency,  in this  paper it  is
referred  to solely  as the  Monoceros  Stream.}. Using  M giant  star
candidates  drawn   from  2MASS,  \citet{2003ApJ...594L.115R}  further
confirmed the existence of the  Monoceros Stream as a distinct stellar
population  beyond the  edge of  the  disk of  the Milky  Way. With  a
galactocentric distance of $18\pm2$kpc,  they find the arc of material
possesses an angular extent of  at $\sim170^o$, with the presence of M
giants  indicating  the  stellar  population of  the  Monoceros  stars
possesses   a    higher   metallicity   than    previously   estimated
$([\frac{Fe}{H}]\sim0.4\pm0.3)$.  \citet{2003ApJ...594L.119C} extended
this  work,   obtaining  velocities  of  2MASS   selected  stars  over
$\sim100^o$  of the  Monoceros  stream.  While  confirming a  velocity
dispersion of $\sim20$km/s, these data  also indicate the stars in the
Monoceros stream orbit  the Galaxy in a prograde  fashion, with little
eccentricity.   While  this  is  somewhat worrisome  for  the  tidally
disrupting dwarf  galaxy hypothesis, \citet{2003ApJ...594L.119C} point
out  that  such features  are  apparent  in  numerical simulations  of
in-plane   dwarf  accretion\citep{2003ApJ...592L..25H}.   Furthermore,
this study  identifies four globular  clusters that are  spatially and
kinematically  aligned  with  the  putative stellar  stream;  such  an
alignment argues against  a Galactic origin for the  stream, such as a
spiral  arm.   Finally,  \citet{2004ApJ...602L..21F} noted  that  five
globular  clusters  aligned with  the  Monoceros  stream,  as well  as
$\sim15$ outer, old stellar clusters that may also be part of the same
population;  these clusters  lie  in a  plane  which is  significantly
tilted  $(\sim17^o)$ to  that of  the  Milky Way.   The population  of
globular clusters is reminiscent of the Sagittarius Dwarf galaxy which
has appears  to have deposited  a similar number of  globular clusters
into  the halo  of the  Milky Way,  bolstering the  argument  that the
Monoceros stream represets a similarly disrupting dwarf galaxy.

The  extensive nature  of  the material  strongly  suggested that  its
origin lay  in a  disrupting system,  in an event  similar to  that of
Sagittarius, but occurring in the  plane of the Galaxy. This, however,
was not  the only  explanation for  its origin as  it may  represent a
previously  unidentified   aspect  of  Galactic   structure.   If  the
disrupting system  hypothesis is correct,  and its destruction  is not
yet  complete,  we should  expect  to  identify  some remnant  of  the
original  dwarf  galaxy.   However,  the expected  location,  observed
against the plane of the Galaxy, makes the detection of such a remnant
difficult.

\section{The Canis Major Dwarf}\label{cmdwarf}
As   with   \citet{2003ApJ...594L.115R},   \citet{2004MNRAS.348...12M}
employed 2MASS to search for a signature of Monoceros stream of stars.
Pushing the search to $|b|\sim5^o$, \citet{2004MNRAS.348...12M} mapped
the density of M-giant stars around the Galactic equator. This reveals
a strong  asymmetry about the  Galactic plane which is  interpreted as
being the Monoceros ring of  stars snaking around the outskirts of the
Milky Way.   Several prominent features are noted,  including a strong
Northern arc  and weaker Southern arc,  both of which  extend for more
than $\sim100^o$ on the sky (see Figure~\ref{fig1}).

Intriguingly,  \citet{2004MNRAS.348...12M} also  identified  a strong,
elliptical  overdensity of  M-stars  at $(l,b)=(240^o,-8^o)$,  aligned
somewhat with  the Galactic disk. The (heliocentric)  distance to this
stellar overdensity is $D_\odot =  7.1\pm0.1$kpc, with a major axis is
$\sim4.2$kpc.   With $\sim2300$  M-giant  stars within  $10^o$ of  its
centre, this overdensity contains a  similar number of M-giants to the
Sagittarius Dwarf  galaxy, a system  which we know is  currently being
cannibalised by the Milky Way.   Given that this implies that the mass
of  the  Canis  Major  overdensity  is  $\sim10^8-10^9{\rm  M_\odot}$,
\citet{2004MNRAS.348...12M} concluded  that it too  represents a dwarf
galaxy also undergoing tidal disruption, and possibly representing the
progenitor of the Monoceros stream of stars.

Unlike Sagittarius, which passes over  the poles of the Milky Way, the
identification of  the Canis Major  dwarf galaxy (as  this overdensity
will  now  be  referred  to)  represents the  first  detection  of  an
accretion occurring within the plane of the Galaxy.

\citet{2004MNRAS.348...12M}  also noted  that  possibly five  globular
clusters  were associated in  phase-space with  the Canis  Major dwarf
galaxy.  This is  a similar number to the  globular cluster population
currently  associated  with the  Sagittarius  dwarf galaxy,  providing
further  evidence  for  the   origin  of  Canis  Major.   Furthermore,
\citet{2004MNRAS.348...12M} noted the phase-space grouping of a number
of four open galactic clusters  directly associated with the main body
of Canis  Major\footnote{Several of these clusters were  also noted by
\citet{2004ApJ...602L..21F} as being part of the Monoceros stream.}.

In  an attempt  to understand  the observed  distribution  of M-stars,
\citet{2004MNRAS.348...12M}  also  undertook  a  series  of  numerical
simulations. Utilizing  detailed models  for the mass  distribution of
the Milky Way  \citep{1998MNRAS.294..429D}, these simulations involved
following the  dynamical dissociation  of dwarf galaxies  (modeled as
King  profiles) as  they  orbited  the Galaxy.   The  results of  this
procedure   favoured  a  dwarf   galaxy  with   an  initial   mass  of
$\sim5\times10^8{\rm   M_\odot}$,    with   a   orbital    period   of
$\sim0.4$Gyrs. While  such simulations can reasonably  account for the
observed distribution of M-stars not  only in the body of Canis Major,
but also along the extensive  arcs above and below the Galactic plane,
the current  data does not allow a  definitive differentiation between
prograde and  retrograde orbits. \citet{2004MNRAS.348...12M}, however,
point out that the resultant orbits of the debris in such an encounter
closely  mimic those  of  stars in  the  thick disk.   Given that  the
estimated  mass of  this single  dwarf  is roughly  $\sim10\%$ of  the
entire thick disk, then the thick  disk could be formed via only a few
such accretion events.

Further observational evidence for the nature of the Canis Major dwarf
came from the study of \citet{aa} who identified the main-sequence and
red giant branch  populations of Canis Major in  the background to the
Galactic open  clusters NGC2477, Tombaugh 1 and  Berkeley 33. Analysis
of   this    population   suggested   it    is   somewhat   metal-rich
$(-0.7<[\frac{Fe}{H}]<0.0)$, with an age of $\sim2-7$Gyrs, although an
apparent blue  plume of stars  is taken as  evidence of a  more recent
episode  of  star formation.   This  study  also  finds a  photometric
parallax  for the  main  body of  Canis  Major of  $\sim8.3\pm1.2$kpc,
larger  than the M-giant  study of  \citep{2004MNRAS.348...12M}, whose
smaller distance determination they  put down to poorer systematics in
the  M-giant photometric  parallax. Finally,  \citet{aa}  also suggest
that as  well as the  identified globular cluster population,  two old
open clusters, AM-2  and Tombaugh 2, are possibly  associated with the
Canis Major dwarf.

Utilizing  the Second  U.S. Naval  Observatory CCD  Astrograph Catalog
(UCAC2),  \citet{bb}  examined the  proper  motions  of  stars in  the
vicinity of  the Canis Major dwarf  galaxy.  Cross-identifying M-giant
stars drawn  from the 2MASS catalog,  the Galactic-longitudinal motion
of  Canis Major  was  found to  be  $\mu_l=-4.0\pm0.4$mas/yr, with  no
measurable motion  in Galactic latitude. At a  distance of $D=8.3$kpc,
this  corresponds to transverse  velocity of  $\sim238\pm28$km/s, with
Canis Major on a prograde orbit  about the Milky Way. Since this value
is  higher than that  predicted by  \citet{2004MNRAS.348...12M}, drawn
from   their   numerical   simulation,    it   is   clear   the   full
three-dimensional  velocity of Canis  Major will  be required  for the
detailed numerical modeling of this accretion event.

Recently, \citet{cc} further  examined the globular cluster population
associated  with  the  Canis  Major  dwarf galaxy,  finding  that  the
age-metallicity  relationship  for these  is  distinct  from the  main
globular  cluster population  of  the Milky  Way.  Furthermore,  these
globular  clusters are  somewhat smaller  than expected  if  they were
drawn from the  Galactic population. Both these lines  of evidence add
further  weight to  the possibility  that the  globular  clusters, and
Canis Major itself, were formed elsewhere, were drawn in presumably as
the  initial orbit  decayed by  dynamical friction,  and  represent an
on-going accretion event onto the Milky Way galaxy.

\section{Further Observations}\label{further}
While the tidally disrupting  dwarf galaxy is the favoured explanation
for  the  observed  overdensity   of  star  in  Canis  Major,  current
observations are  not yet  completely conclusive, and  the possibility
that  the  Monoceros  Ring  and  Canis  major  dwarf  galaxy  actually
represent some unknown aspect of Galactic structure has not been ruled
out.  Hence,  several observational  programs are underway  to address
this issue.

The    first    is    an    extension    of    the    earlier    study
by~\citet{2003MNRAS.340L..21I}, systematically mapping above and below
the Galactic plane with  wide-field camera observations. This has been
completed  in the  north, using  the  Wide-Field Camera  on the  Isaac
Newton Telescope; these data are  being analyzed and will be published
shortly (Conn et al. {\it  in preparation}). The southern survey, with
the Wide-Field Imager at  the Anglo-Australian Telescope, is currently
underway  and  will be  completed  by  mid-2004.   These will  provide
important probes  of the  extent of the  stellar material, as  well as
constraining its distance (via main sequence fitting).

\begin{figure}
\includegraphics[height=4.7in]{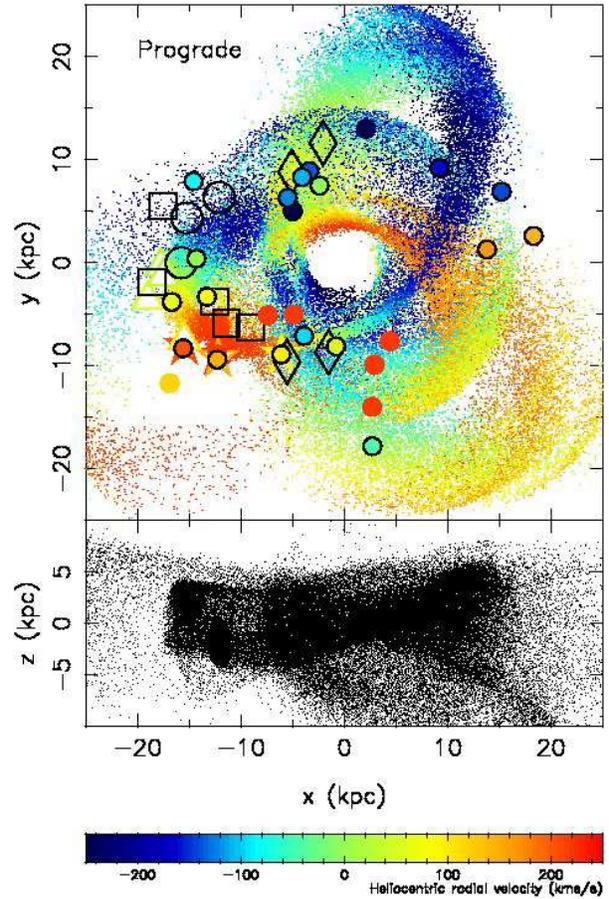}
\caption{\label{fig2}  A numerical  simulation  of the  demise of  the
Canis Major dwarf  galaxy. The top panel presents  the view from above
the Milky  Way [with  the Sun located  at $(x,y,z)=-8,0,0\  $kpc]. The
lower panel presents the side-on view of the debris. The colour coding
in both denoted the heliocentric velocity of the stars (as given in the
lower key), while  the symbols represent data from  the INT survey and
2MASS [Figure 14 from \citet{2004MNRAS.348...12M}].}
\end{figure}

To augment these  studies, 2dF observations of the  main body of Canis
Major and the extensive stream  of stars are being conducted to obtain
stellar kinematics,  via the calcium triplet, over  $\sim100^o$ of the
sky.   Coupled with  the  spatial data  obtained  with the  wide-field
camera surveys, these data should test the hypotheses that Canis Major
represents a  truly disrupting dwarf galaxy,  or (equally interesting)
is a  currently unknown aspect of Galactic  structure. Furthermore, if
the dwarf galaxy is confirmed,  then these data will provide important
constraints to numerical simulations of its orbit and eventual demise.

\section*{Acknowlegements}
GFL  thanks Joss  Bland-Hawthorn for  inviting him  to the  Little Bay
Meeting.  The  anonymous referee  is  thanked  for their  constructive
comments.

 \label{lastpage}


\begin{thebibliography}{}
%
\bibitem[\protect\citeauthoryear{Bellazzini     et     al.}{2004}]{aa}
Bellazzini  M., Ibata  R., Monaco  L., Martin  N., Irwin  M.~J., Lewis
G.~F., 2004, MNRAS in press, {\it astro-ph/0311119}
%
\bibitem[\protect\citeauthoryear{Brook                               et
al.}{2003}]{2003ApJ...585L.125B} Brook C.~B., Kawata D., Gibson B.~K.,
Flynn C., 2003, ApJ, 585, L125
%
\bibitem[\protect\citeauthoryear{Crane et al.}{2003}]{2003ApJ...594L.119C} 
Crane J.~D., Majewski S.~R., Rocha-Pinto H.~J., Frinchaboy P.~M., Skrutskie 
M.~F., Law D.~R., 2003, ApJ, 594, L119 
%
\bibitem[\protect\citeauthoryear{Dehnen \& 
Binney}{1998}]{1998MNRAS.294..429D} Dehnen W., Binney J., 1998, MNRAS, 294, 
429 
%
\bibitem[\protect\citeauthoryear{Forbes et al.}{2004}]{cc} 
Forbes, D.~A., Strader, J., Brodie, J.P., 2004, {\it astro-ph/0403136}
%
\bibitem[\protect\citeauthoryear{Frinchaboy et 
al.}{2004}]{2004ApJ...602L..21F} Frinchaboy P.~M., Majewski S.~R., Crane 
J.~D., Reid I.~N., Rocha-Pinto H.~J., Phelps R.~L., Patterson R.~J., Mu{\~ 
n}oz R.~R., 2004, ApJ, 602, L21 
%
\bibitem[\protect\citeauthoryear{Helmi et al.}{2003}]{2003ApJ...592L..25H} 
Helmi A., Navarro J.~F., Meza A., Steinmetz M., Eke V.~R., 2003, ApJ, 592, 
L25 
%
\bibitem[\protect\citeauthoryear{Helmi et al.}{1999}]{1999Natur.402...53H} 
Helmi A., White S.~D.~M., de Zeeuw P.~T., Zhao H., 1999, Natur, 402, 53 
%
\bibitem[\protect\citeauthoryear{Ibata et al.}{2003}]{2003MNRAS.340L..21I} 
Ibata R.~A., Irwin M.~J., Lewis G.~F., Ferguson A.~M.~N., Tanvir N., 2003, 
MNRAS, 340, L21 
%
\bibitem[\protect\citeauthoryear{Ibata et al.}{2004}]{bb} 
Ibata R., Bellazzini, M., Irwin M., Lewis G.~F., Martin, N.~F., 
2004, MNRAS (Submitted)
%
\bibitem[\protect\citeauthoryear{Ibata et al.}{2002}]{2002MNRAS.332..921I} 
Ibata R.~A., Lewis G.~F., Irwin M.~J., Cambr{\' e}sy L., 2002, MNRAS, 332, 
921 
%
\bibitem[\protect\citeauthoryear{Ibata et al.}{2001}]{2001ApJ...551..294I} 
Ibata R., Lewis G.~F., Irwin M., Totten E., Quinn T., 2001, ApJ, 551, 294 
%
\bibitem[\protect\citeauthoryear{Klypin et al.}{1999}]{1999ApJ...522...82K} 
Klypin A., Kravtsov A.~V., Valenzuela O., Prada F., 1999, ApJ, 522, 82 
%
\bibitem[\protect\citeauthoryear{Majewski et 
al.}{2003}]{2003ApJ...599.1082M} Majewski S.~R., Skrutskie M.~F., Weinberg 
M.~D., Ostheimer J.~C., 2003, ApJ, 599, 1082 
%
\bibitem[\protect\citeauthoryear{Martin et al.}{2004}]{2004MNRAS.348...12M} 
Martin N.~F., Ibata R.~A., Bellazzini M., Irwin M.~J., Lewis G.~F., Dehnen 
W., 2004, MNRAS, 348, 12 
%
\bibitem[\protect\citeauthoryear{Newberg et al.}{2002}]{2002ApJ...569..245N} 
Newberg H.~J.~et al., 2002, ApJ, 569, 245 
%
\bibitem[\protect\citeauthoryear{Rocha-Pinto et 
al.}{2003}]{2003ApJ...594L.115R} Rocha-Pinto H.~J., Majewski S.~R., 
Skrutskie M.~F., Crane J.~D., 2003, ApJ, 594, L115 
%
\bibitem[\protect\citeauthoryear{Yanny et al.}{2003}]{2003ApJ...588..824Y} 
Yanny B.~et al., 2003, ApJ, 588, 824 
%
\bibitem[Yanny et al.(2004)]{2004ApJ...605..575Y} Yanny, B., et al.\ 2004, 
ApJ, 605, 575 
%
\end{thebibliography}
\end{document}